\documentclass[11pt]{article}
\usepackage{moriond}

\def\Journal#1#2#3#4{{#1} {\bf #2}, #3 (#4)}

\def\PLB{{\em Phys. Lett.}  B}
\def\PRL{\em Phys. Rev. Lett.}
\def\PRD{{\em Phys. Rev.} D}

\def\be{\begin{equation}}
\def\ee{\end{equation}}
\def\bea{\begin{eqnarray}}
\def\eea{\end{eqnarray}}

\newcommand{\Photo}{}

\begin{document}
\vspace*{4cm}
\title{MEASUREMENT OF HEAVY-FLAVOR PROPERTIES AT CMS AND ATLAS}

\author{ R. COVARELLI }

\address{Department of Physics and Astronomy, University of Rochester, 14627, Rochester, NY, United States}

\maketitle\abstracts{
Thanks to the excellent performances of ATLAS and CMS in triggering on muon 
signals and reconstructing these particles down to low transverse momentum,
large samples of heavy-flavored hadrons have been collected 
in the 2011 LHC run at $\sqrt{s} = 7$ TeV. The analysis of these samples has
enabled both experiments to perform competitive measurements of heavy-flavor
properties, such as quarkonium polarization, lifetime and $CP$-violation
measurements, hadron spectroscopy and branching ratios of rare $B$ decays.} 

ATLAS and CMS capabilities in heavy-flavor physics are almost entirely based
on muon and 
di-muon triggers, which can collect signals down to low transverse momenta
(typical values for thresholds range between 3 and 6 GeV/$c$), while
maintaining a reasonable background rate by using high-level selection
criteria. Muons are reconstructed offline
using techniques which match information from the inner tracking detectors
and the muon chambers, and are combined to reconstructed charged tracks to
form partially and fully reconstructed final states, like 
for instance $B_s \rightarrow J/\psi \phi$. 
All analyses presented are based on the data collected
by ATLAS and CMS in 2011, at a center-of-mass energy of 7 TeV, which amounts
to an integrated luminosity up to 5.0 fb$^{-1}$.

\section{Quarkonium Polarization}

Theoretical predictions on quarkonium polarization at production in hadron 
colliders are still controversial. Calculations in Non-Relativistic QCD (NRQCD)
schemes, including the color-octet mechanism, require inputs from experimental 
data~\cite{butkni} and the effect of feed-down from heavier charmonium
states can be significant~\cite{shaochao}. Recent results from the
CDF experiment~\cite{cdfpol}, which use advanced experimental techniques,
including a simultaneous determination of all polarization parameters in
difference reference frames~\cite{lamerda}, indicate that $\Upsilon$ 
mesons are not significantly polarized at high transverse momentum 
($p_T \gg m_{\Upsilon}$). NRQCD would predict preponderance of transverse
polarization.

\begin{figure}
\centerline{\includegraphics[width=0.8\linewidth]{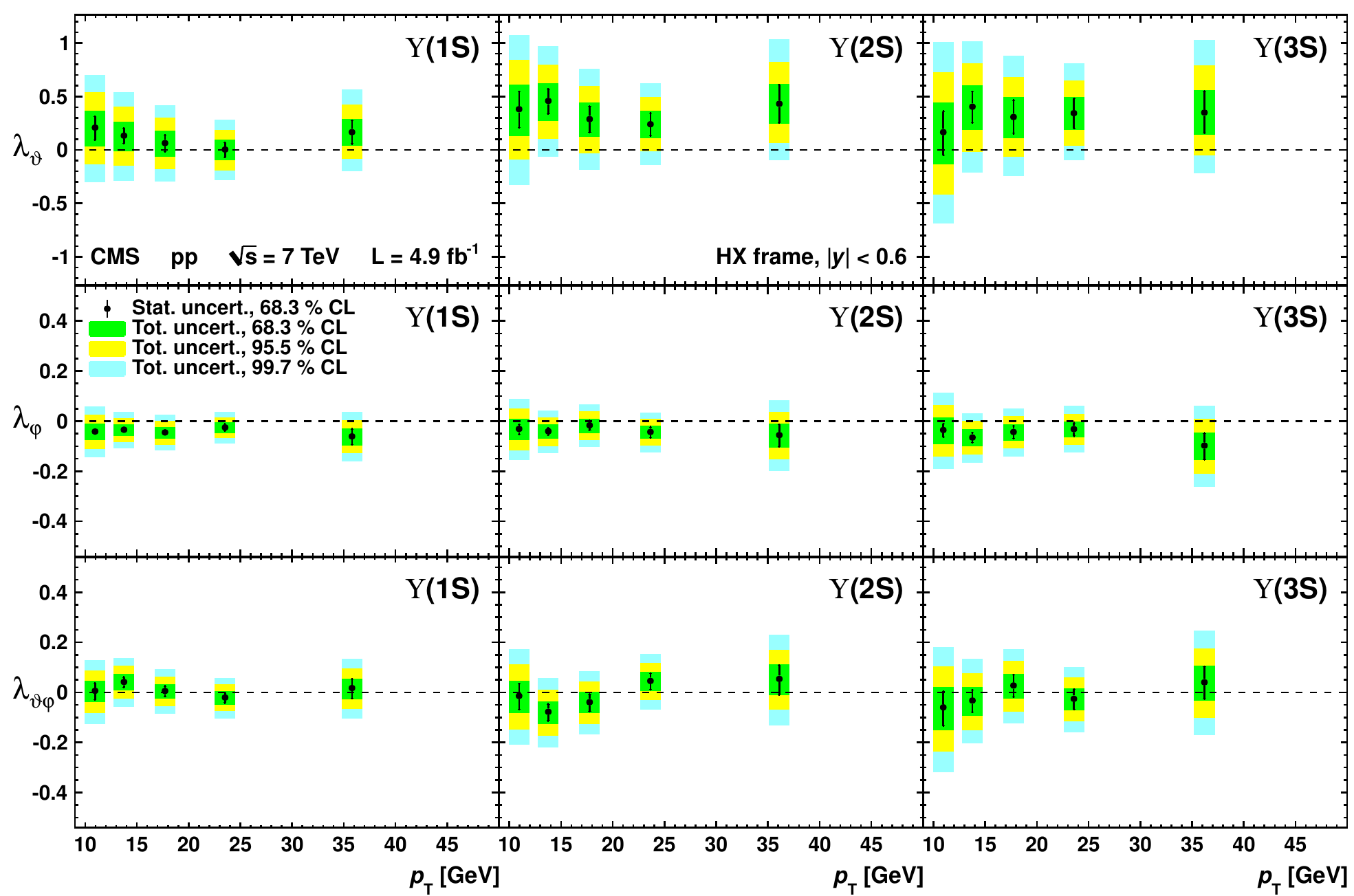}}
\caption{Fitted values of the polarization parameters $\lambda_\theta$, $\lambda_\phi$, $\lambda_{\theta\phi}$ (rows) in the helicity frame and in the
rapidity bin $|y| < 0.6$ for the three states $\Upsilon(1\mathrm{S})$,
$\Upsilon(2\mathrm{S})$ and $\Upsilon(3\mathrm{S})$ (columns). The notation for
uncertainties is explained in the legend.}
\label{fig:pol}
\end{figure}

CMS~\cite{cmspol} performed a measurement of $\Upsilon(1\mathrm{S})$,
$\Upsilon(2\mathrm{S})$ and $\Upsilon(3\mathrm{S})$ polarizations in a $p_T$
range of 10-50 GeV/$c$ and two rapidity bins. The method is based on the
construction of posterior probability distributions which depend on the 
values of the polarization angles $\cos\theta$ and $\phi$ of 
reconstructed dimuons and on the 
reconstruction efficiencies, computed as a function of muon kinematic 
variables. Background is subtracted using fits to the dimuon invariant mass
distributions. Muon efficiencies are extracted from data-driven 
methods~\cite{muo004}. The result is extracted in several reference frames
and the result is cross-checked using frame-independent polarization
parameters. Results for the three quarkonium states in the helicity frame 
are shown
in Fig.~\ref{fig:pol}: all of them are compatible with zero within the total
uncertainties.
 
\section{Lifetimes and $CP$ violation}

$B$-hadron lifetime values (and lifetime differences in case of 
oscillating meson systems) represent important tests of Heavy Quark Effective Theory
(HQET) and lattice QCD. Experimentally the least known among $B$-hadron 
lifetimes is
$\tau(\Lambda_b)$: CDF and D0 results are not in agreement~\cite{tevtau} on its
measured value and the PDG average
value shows a discrepancy with HQET calculations up to order $1/m_b^2$, therefore
specific corrections must be introduced to improve the consistency~\cite{gabbiani}. 

ATLAS and CMS~\cite{lhctau} have measured the $\Lambda_b$ lifetime, with ATLAS
also reporting a mass measurement. In both experiments the decay channel
$J/\psi \Lambda^0 \rightarrow \mu^+ \mu^- p \pi^-$ has been used to reconstruct
the $\Lambda_b$ baryon, profiting from the clean dimuon signature and the displaced decay
vertex of the $\Lambda^0$. The mass and lifetime values have been extracted from simultaneous fits to the $J/\psi \Lambda^0$ invariant mass (using mass constraints
of the sub-products) and proper decay time distributions. Selection
biases have been corrected using simulation. ATLAS obtains: $m(\Lambda_b) = 5619.7 \pm 0.7_{\mathrm{stat.}} \pm 1.1_{\mathrm{syst.}}$ MeV/$c^2$ and
$\tau(\Lambda_b) = 1.449 \pm 0.036_{\mathrm{stat.}} \pm 0.017_{\mathrm{syst.}}$
ps, while CMS measures $\tau(\Lambda_b) = 1.503 \pm 0.052_{\mathrm{stat.}} \pm 0.031_{\mathrm{syst.}}$ ps. Both results are in better agreement with CDF 
than with D0 results, although the compatibility between the ATLAS and D0
values is within 1.6$\sigma$.

Of particular relevance for the indirect search for New Physics (NP) in 
processes described by ``box'' and ``loop'' diagrams is the measurement of 
the $B_s$ mixing phase ($\phi_s = -2 \beta_s$), whose value in the Standard
Model is tightly constrained by precise measurements of quantities related 
to the unitarity triangle in the $B_d$ system~\cite{utfit}. Its value, 
computed in the Standard Model (SM) hypothesis (only one $CP$-violating phase) is 
$\phi_s = -0.0364 \pm 0.0016$ rad. Experimentally the measurement can be performed using  
$B_s$ decays to $CP$ eigenstates. In the results presented here, ATLAS and CMS do not use techniques to identify
the flavor of the $B_s$ meson at production and the sensitivity to $\phi_s$
remains limited. CMS only measured $\Delta\Gamma_s$, fixing $\phi_s$ to the
SM value, while ATLAS has determined both. 

\begin{figure}
\centerline{\includegraphics[height=0.35\linewidth]{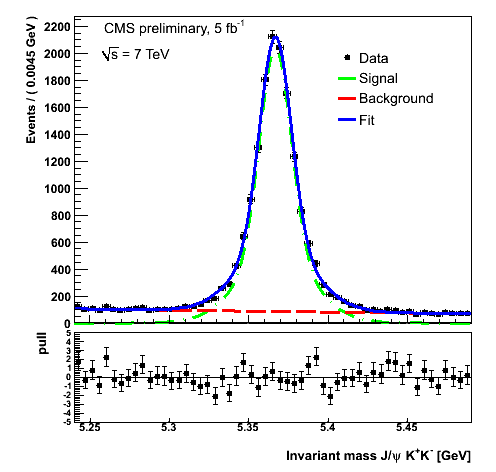}
\includegraphics[height=0.35\linewidth]{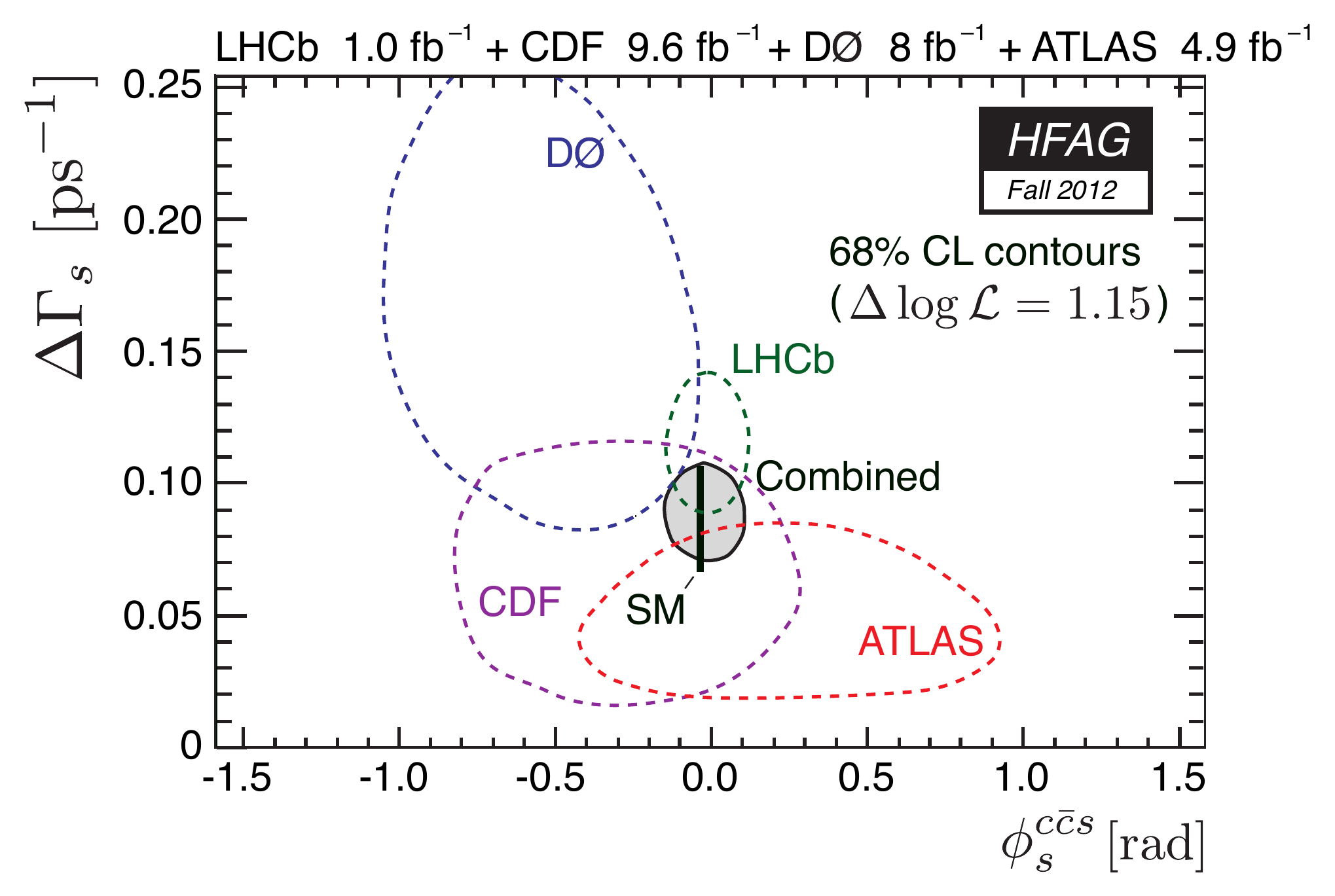}}
\caption{Left: Reconstructed $J/\psi \phi$ invariant mass in CMS with projections
of the 5-dimensional fit: colored lines are explained in the legend. Right:
Combination of ATLAS, Tevatron and LHCb results in the 
$\Delta\Gamma_s$-$\phi_s$ plane.}
\label{fig:psiphi}
\end{figure}

ATLAS and CMS~\cite{lhcpsiphi} use the $B_s \rightarrow J/\psi \phi \rightarrow \mu^+ \mu^- K^+ K^-$ final state, which is relatively abundant and clean 
(Fig.~\ref{fig:psiphi} left). Because of the entangled $CP$-even and -odd 
contributions in the vector-vector meson system, the result is obtained from
a five-dimensional fit to $B_s$ mass, proper decay time, and three angular
variables related to the distribution in space of the final decay products.
The CMS result is $\Delta\Gamma_s = 0.048 \pm 0.024_{\mathrm{stat.}} \pm 0.003_{\mathrm{syst.}}$ ps$^{-1}$, while ATLAS measures $\Delta\Gamma_s = 0.053 \pm 0.021_{\mathrm{stat.}} \pm 0.010_{\mathrm{syst.}}$ ps$^{-1}$ and $\phi_s = 0.22 \pm 0.41_{\mathrm{stat.}} \pm 0.10_{\mathrm{syst.}}$. While both experiments report values of $\Delta\Gamma_s$ noticeably smaller
than the latest high-precision measurement by LHCb~\cite{lhcbpsiphi},
this discrepancy is not statistically significant. Fig.~\ref{fig:psiphi} right
shows the comparison of ATLAS and LHCb results using two-dimensional
likelihood contours~\cite{hfag}.

\section{Hadron spectroscopy}

Theories which explain the formation of hadrons often predict a very large
number of states with similar quark structure and different quantum numbers.
The quark model predicts baryonic combinations with one or more
bottom and charm quarks, many of which have not been observed yet.
Similarly, the spectroscopy of $c{\bar c}$, $b{\bar b}$ and $c{\bar b}$ states 
has been experimentally confirmed in many cases. Still states exist 
(even below the open-charm and -bottom thresholds) which have
never been identified. On the other hand, the existence of a 
few ``unconventional'' states, like the $X(3872)$, which are interpreted as
bound states but do not fit in the 
spectroscopy predicted by potential models, has been established. 
 
\begin{figure}
\centerline{\includegraphics[height=0.35\linewidth]{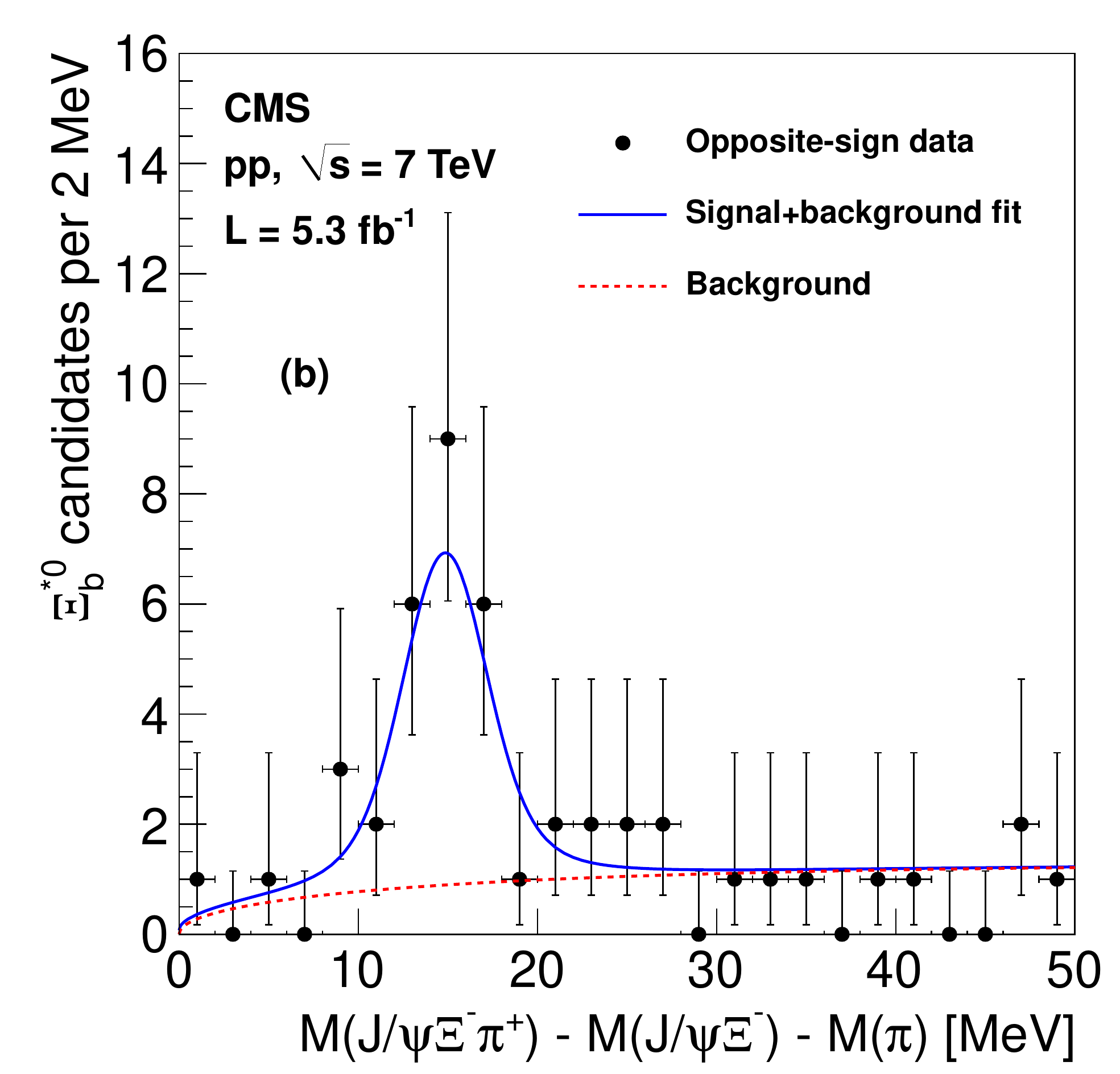}
\includegraphics[height=0.35\linewidth]{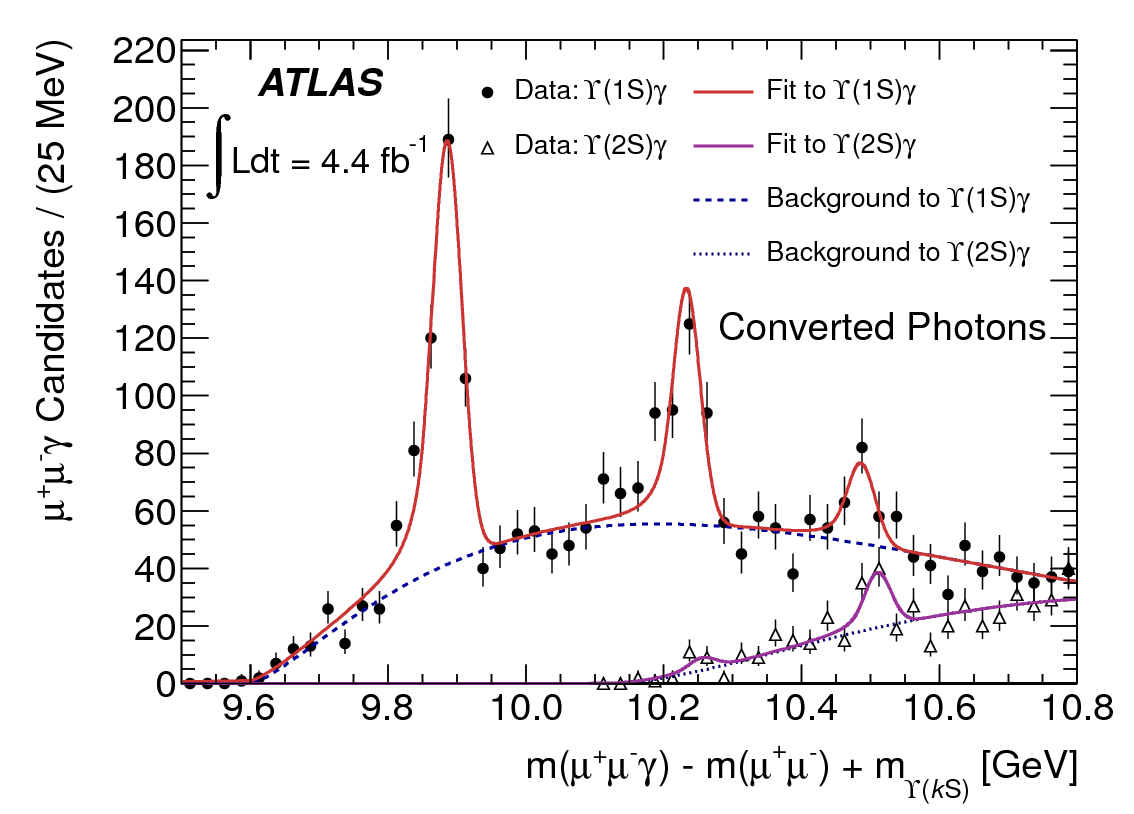}}
\caption{Left: Invariant mass difference ($\Delta m$) between the reconstructed 
$\pi^+ \Xi_b^-$ and $\Xi_b^-$ systems in CMS. About 20 events are peaking at
$\Delta m \simeq 15$ MeV/$c^2$, indicating a resonant state.
Right: $\Upsilon(n\mathrm{S}) \gamma$ ($n = 1,2$) reconstructed invariant mass in ATLAS, fixing the $\Upsilon(n\mathrm{S})$ masses to their PDG values. The rightmost peaks correspond to the newly-observed
$\chi_b(3\mathrm{P})$ state.}
\label{fig:bary}
\end{figure}

ATLAS and CMS have performed searches for conventional and exotic hadron 
states. Using the decay chain $\Xi_b^{0*} \rightarrow  \pi^+ \Xi_b^- 
\rightarrow \pi^+ J/\psi \Xi^- \rightarrow \pi^+ \mu^+ \mu^- \pi^- \Lambda^0
\rightarrow \pi^+ \mu^+ \mu^- \pi^- p \pi^-$, CMS has observed the $J^P
= 3/2^+$ partner of the $\Xi_b^{0}$. The presence of three weakly decaying
particles in the chain, giving rise to detached vertices, has allowed to
isolate a clean signal, as shown in Fig.~\ref{fig:bary} left. 
The measured mass is $5945.0 \pm 0.7_{\mathrm{stat.}} \pm 0.3_{\mathrm{syst.}} \pm 2.7_{m(\Xi_b^-)}$ MeV/$c^2$, in agreement with
theoretical predictions~\cite{ernestion}. ATLAS reported the first observation of the
$\chi_b(3\mathrm{P})$ bottomonium state (actually the superposition of three
states with different total spin) using the decay mode $\chi_b(3\mathrm{P})
\rightarrow \Upsilon(1\mathrm{S}, 2\mathrm{S}) \gamma \rightarrow \mu^+ \mu^- \gamma$ or $\rightarrow \mu^+ \mu^- e^+ e^-$, where the photon conversion occurs inside the detector material~\cite{atlasion}. The signals are shown in Fig.~\ref{fig:bary} right (for the converted-photon case), and the 
measured mass is $10530 \pm 5_{\mathrm{stat.}} \pm 9_{\mathrm{syst.}}$ MeV/$c^2$. Both observations exceed 6$\sigma$ significance.

\begin{figure}
\centerline{\includegraphics[angle=90,height=0.35\linewidth]{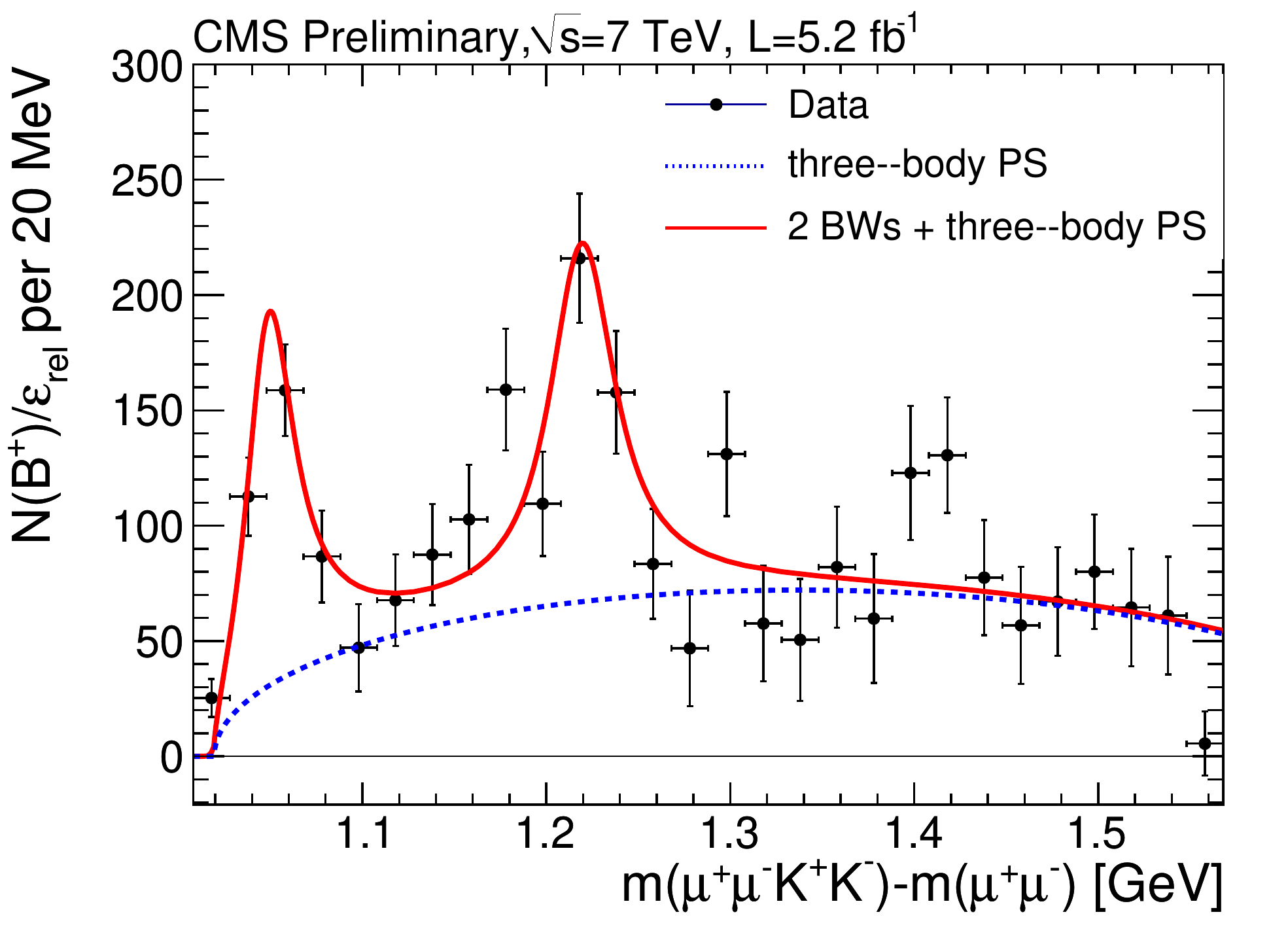}}
\caption{Invariant mass difference ($\Delta m$) between the reconstructed 
$J/\psi \phi$ and $J/\psi$ systems in CMS, obtained with a bin-by-bin
background subtraction technique. The distribution is fitted with a 
phase-space background and two Breit-Wigner shapes convoluted with
resolution functions obtained from simulation.}
\label{fig:kai}
\end{figure}

CMS and ATLAS have detected clear signals of $B_c^{\pm} \rightarrow J/\psi 
\pi^{\pm}$ (CMS also reporting evidence for $B_c^{\pm} \rightarrow J/\psi \pi^+ \pi^- \pi^{\pm}$) from
which cross sections and ratios of branching fractions will be measured.
CMS has also investigated possible resonant structures of the $J/\psi \phi$
system in the $B^{\pm} \rightarrow J/\psi \phi K^{\pm}$ decay, in order to
confirm the CDF observation of a structure at $m_{J/\psi \phi} \simeq 4140$
MeV/$c^2$~\cite{kaiyi}. As shown in Fig.~\ref{fig:kai},
two structures are clearly seen in the bin-by-bin subtracted invariant mass
spectrum of the $J/\psi \phi$
system. Further studies are ongoing in order to understand the exact nature
of these structure.   

\section{Rare $B$ decays}

In ATLAS and CMS searches for $B_{d,s} \rightarrow \mu^+ \mu^-$ decays
have been performed~\cite{lhcbmumu}. These decays are predicted to be rare in the SM and 
a significant enhancement over 
the SM branching ratios ($\mathcal{B}_{\mathrm{SM}} \sim 10^{-10}-10^{-9}$~\cite{burasse})
is possible
in most supersymmetric theories~\cite{paride}. An observation of a non-SM value of the
branching fraction would therefore represent an indirect evidence for NP.

In both experiments, a ``normalization'' sample of events with $B^+ \rightarrow J/\psi K^+$ decays 
is used 
to remove uncertainties related to the $b{\bar b}$ production cross section 
and the integrated luminosity, and to reduce uncertainties on efficiencies, which 
are determined from simulation. Selection is based on several variables,
including high $p_T$, vertex displacement and dimuon isolation, which have 
been chosen to mitigate the effects of high pileup. 
A ``blind'' analysis approach is applied.  
In ATLAS the selection variables are included in a Boosted Decision
Tree (BDT), while CMS uses a cut-based selection, and both analyses are
optimized for the best upper limit.
In CMS,
because of better mass resolution, the signal is separated in $B_d$ and $B_s$
regions.

Event-counting experiments are performed in dimuon mass regions around the 
$B_s$ and $B_d$ masses. 
Monte Carlo simulations are used to estimate backgrounds due to other rare 
$B$ decays and 
combinatorial backgrounds are evaluated from the data in dimuon invariant mass 
sidebands. In all cases, the observed number of events is consistent with 
background plus SM signals.
In CMS the resulting upper limits on the branching fractions are  
${\cal B}(B_s \rightarrow \mu^+ \mu^-) < 7.7\times10^{-9}$ and ${\cal B}(B_d \rightarrow \mu^+ \mu^-) < 1.8\times10^{-9}$
at 95\% CL, while ATLAS sets the limit ${\cal B}(B_s \rightarrow \mu^+ \mu^-) < 2.2\times10^{-8}$ at 95\% CL. It has been predicted~\cite{aykroyd} by naive luminosity scaling that 
CMS is expected to observe a SM ${\cal B}(B_s \rightarrow \mu^+ \mu^-)$ with
more than 3$\sigma$ significance using the full data set at $\sqrt{s} = 8$
TeV.


\section*{References}

\begin{thebibliography}{99}

\bibitem{butkni} M. Butensch\"{o}n and B.A. Kniehl, \Journal{{\it Mod. Phys. Lett.} A}{28}{1350027}{2013}.

\bibitem{shaochao} H.S. Shao and K.T.Chao, \texttt{arXiv:1209.4610}.

\bibitem{cdfpol} CDF Collaboration, \Journal{\PRL}{108}{151802}{2012}.

\bibitem{lamerda} P. Faccioli {\it et al.}, \Journal{\PRL}{105}{061601}{2010}.

\bibitem{cmspol} CMS Collaboration, \Journal{\PRL}{110}{081802}{2012}.

\bibitem{muo004} CMS Collaboration, \Journal{{\it JINST}}{7}{P10002}{2012}.

\bibitem{tevtau} CDF Collaboration, \Journal{\PRL}{106}{121804}{2011}; D0 Collaboration, \Journal{\PRD}{85}{112003}{2012}.

\bibitem{gabbiani} F. Gabbiani {\it et al.}, \Journal{\PRD}{70}{094031}{2004}.

\bibitem{lhctau} CMS Collaboration, \texttt{arXiv:1304.7495}, submitted to {\it JHEP} (2013); ATLAS Collaboration, \Journal{\PRD}{87}{032002}{2012}.

\bibitem{utfit} \texttt{http://ckmfitter.in2p3.fr/www/results/plots\_ichep12/num/ckmEval\_results.html}.

\bibitem{lhcpsiphi} CMS Collaboration, CMS-PAS-BPH-11-006, ATLAS Collaboration \Journal{{\it JHEP}}{12}{072}{2012}.

\bibitem{lhcbpsiphi} LHCb Collaboration, \texttt{arXiv:1304.2600}, submitted to \PRD~(2013).

\bibitem{hfag} \texttt{http://www.slac.stanford.edu/xorg/hfag/osc/fall\_2012}.

\bibitem{ernestion} CMS Collaboration, \Journal{\PRL}{108}{252002}{2012}.

\bibitem{atlasion} ATLAS Collaboration, \Journal{\PRL}{108}{152001}{2012}.

\bibitem{kaiyi} CDF Collaboration, \Journal{\PRL}{102}{242002}{2009};

\bibitem{lhcbmumu} CMS Collaboration, \Journal{{\it JHEP}}{04}{033}{2012}; ATLAS Collaboration, \Journal{\PLB}{713}{180}{2012}.

\bibitem{burasse} A. Buras {\it et al.}, \Journal{{\em Eur. Phys. J.} C}{72}{2172}{2012}.

\bibitem{paride} G. Isidori and P. Paradisi, \Journal{\PLB}{639}{499}{2006}.

\bibitem{aykroyd} D. Akeroyd {\it et al.}, \Journal{{\it JHEP}}{12}{088}{2011}.

\end{thebibliography}
\end{document}